\documentclass[aps,prb,superscriptaddress]{revtex4}
\usepackage{graphicx}
\usepackage{amsfonts}
\usepackage{amsmath}
\usepackage{amssymb}

\def\sgn{{\rm sgn}}

\begin{document}

\title{Josephson current in a superconductor-ferromagnet junction with
two noncollinear magnetic domains}
\author{B.\ Crouzy}
\affiliation{Institute for Theoretical Physics,
\'Ecole Polytechnique F\'ed\'erale de Lausanne (EPFL), 
CH-1015 Lausanne, Switzerland}
\author{S.\ Tollis}
\affiliation{Institute for Theoretical Physics,
\'Ecole Polytechnique F\'ed\'erale de Lausanne (EPFL), 
CH-1015 Lausanne, Switzerland}
\author{D.\ A.\ Ivanov}
\affiliation{Institute for Theoretical Physics,
\'Ecole Polytechnique F\'ed\'erale de Lausanne (EPFL), 
CH-1015 Lausanne, Switzerland}

\date{\today}
\begin{abstract}
We study the Josephson effect in a superconductor--ferromagnet--superconductor (SFS)
junction with ferromagnetic domains of noncollinear magnetization. As a model for
our study we consider a diffusive junction with two ferromagnetic domains along
the junction. The superconductor is assumed to be close to the critical temperature $T_c$,
and the linearized Usadel equations predict a sinusoidal current-phase relation. We find
analytically the critical current as a function of domain lengths and of the angle
between the orientations of their magnetizations. As a function of those parameters,
the junction may undergo transitions between $0$ and $\pi$ phases. We find that
the presence of domains reduces the range of junction lengths at which the $\pi$
phase is observed. For the junction with two domains of the same
length, the $\pi$ phase totally disappears as soon as the misorientation angle exceeds $\frac{\pi}{2}$. We further comment
on the possible implication of our results for experimentally observable $0$--$\pi$
transitions in SFS junctions.
\end{abstract}

\maketitle

\section{Introduction}

The interest in proximity structures made of superconducting and
ferromagnetic layers (respectively, S and F) in contact with each other has
been recently renewed due their potential applications to spintronics 
\cite{spintronics} and to quantum computing.\cite{quantum,dima} 
The interplay between superconductivity (which tends to organize the electron 
gas in Cooper pairs with opposite spins) and ferromagnetism (which tends to align spins 
and thus to destroy the Cooper pairs) leads to a variety of surprising 
physical effects (for a review, see Ref.\ \onlinecite{Rev}). 
As a consequence of the exchange splitting of the Fermi level,\cite{demler} the Cooper pair
wave function shows damped oscillations in the ferromagnet, leading to the appearance of the
so-called ``$\pi$ state'' in SFS junctions.\cite{bula} In
the $\pi$ state, the superconducting order parameter is of opposite
sign in the two S electrodes of the Josephson junction, and thus a macroscopic 
superconducting phase difference of $\pi$ appears in the thermodynamic equilibrium. 
This phase difference should lead to spontaneous nondissipative
currents in a Josephson junction with annular geometry.\cite{annular}
A possible signature for the $\pi$-state appearance is a cancellation of the
Josephson critical current followed by a reversal of its sign as a function of the
junction length.\cite{Rev}
The recent experimental observations of critical-current oscillations in 
experiments \cite{kontos,pi1st,pi2nd,Dead} have demonstrated such $0$--$\pi$ transitions
as a function of the ferromagnet thickness and temperature.

The appropriate formalism to deal with mesoscopic S/F junctions has been derived by 
Eilenberger.\cite{eilenberger} The equations of motion for the quasiclassical Green function 
(averaged over the fast Fermi oscillations) can be further simplified 
in the diffusive regime, \textit{i.e.}, when the motion of the electrons is governed 
by frequent scattering on impurity atoms: the Green functions can then be averaged 
over the momentum directions. This averaging is justified as long as the elastic mean 
free path $l_{e}$ is much smaller than the relevant length scales of the system, 
namely the size of the layers, the
superconducting coherence length $\xi_{S}=\sqrt{D/2\pi{T_{c}}}$, and
the length characterizing the Cooper pair wave function decay in the
ferromagnet $\xi_{F}=\sqrt{D/h}$. Here and in the following, $D$
denotes the diffusion constant, $T_{c}$ the superconducting critical
temperature, $h$ the magnitude of the exchange field, and the system of units
with $\hbar=k_{B}=\mu_{B}=1$ is chosen.
The diffusive limit is reached in most of the experimental realizations of S/F
heterostructures. In this limit, the Green functions can be combined in a
4$\times$4 matrix in the Nambu $\otimes$ spin space, and this matrix
obeys the Usadel equations.\cite{Us} SFS Josephson junctions with homogeneous
magnetization have been studied in detail within this framework.\cite{Rev} 
Close to $T_{c}$, the proximity effect is weak, the Usadel equations can be
linearized,  and the current-phase relation is sinusoidal\cite{golubov}
\begin{equation}  \label{cpr}
I_J=I_{c}\sin \phi \, ,
\end{equation}
where $\phi$ is the superconducting phase difference across the junction.
The critical current $I_{c}$ shows a damped oscillatory dependence 
on the F-layer thickness (for a review, see Refs.\ \onlinecite{buzdin82} 
and\ \onlinecite{buzdin91}).
 
However, understanding the effect of a nonhomogeneous magnetization 
is of crucial interest for obtaining a good quantitative description 
for the critical-current oscillations in SFS junctions. 
Indeed it is known that real ferromagnetic compounds usually 
have a complex domain structure. Strong ferromagnets (such as Ni or Fe) 
consist of domains with homogeneous magnetization pointing in different 
directions whereas the magnetic structure of the weak ferromagnets 
(Cu-Ni and Pd-Ni alloys) used in the experiments reported in 
Refs.\ \onlinecite{kontos,pi1st,pi2nd,Dead}
is still not known precisely. The problem of SFS junctions with
inhomogeneous magnetization has been addressed previously
for spiral magnetizations\cite{Bergeret} and in the
case of domains with antiparallel (AP) magnetizations.\cite{Hekking,Koshina2} 
In the latter case, the critical-current oscillations (and thus the $\pi$ state) 
are suppressed in the symmetric case where the F layer consists of two domains of the same size.
This can be explained by a compensation between the phases acquired by the 
Andreev reflected electrons and holes, of opposite spins, in the two domains.\cite{Hekking}

In the present paper, we extend that analysis to the case of a SFF$'$S
junction close to $T_{c}$, with the two magnetic domains F and F$'$ of
arbitrary length and relative orientation of the magnetizations. 
To emphasize the effect of the misorientation angle between the magnetizations of the
two domains, we choose to minimize the number of parameters in the model. 
The interfaces are then chosen to be perfectly transparent, and spin-flip scattering
is neglected in both S and F layers. 
Furthermore, we assume that the diffusive limit is fully reached, 
that is we do not take into account corrections due to a finite mean free path 
(note that for strong ferromagnets the magnetic coherence length $\xi_{F}$ may become 
comparable to $l_{e}$). 

The main result of our calculation is that, in the symmetric case where the two
domains have equal thicknesses, we obtain a progressive reduction of the 
$\pi$-state region of the phase diagram as the misorientation angle increases.
Surprisingly, the $\pi$ state completely
disappears as soon as the misorientation angle $\theta$ exceeds $\frac{\pi}{2}$. 

The paper is organized as follows. In Section II we solve the linearized Usadel
equations and give the general expression for the Josephson current. In Section III 
we discuss the simplest cases of parallel and antiparallel relative orientation 
of magnetizations with different domain sizes $d_{1}$ and $d_{2}$. 
We obtain analytically the full phase diagram in $d_{1}$--$d_{2}$ coordinates. 
In agreement with Ref.\ \onlinecite{Hekking}, the $\pi$ state is absent in the
symmetric case $d_{1}=d_{2}$ for domains with antiparallel magnetizations. 
In the asymmetric case,
the critical current oscillates as $d_{2}$ is varied while keeping $d_{1}$
constant. For sufficiently thick layers ($d_{1,2}\gg \xi_F$), 
the critical-current oscillations behave like in a single domain of
thickness $|d_{1}-d_{2}|+(\pi/4)\xi_F$. In Section IV, we discuss the case of an arbitrary
misorientation angle in the symmetric configuration $d_{1}=d_{2}=d$.
In the limit when the exchange field is much larger than $T_{c}$, we derive
analytically the $0$--$\pi$ phase diagram of the junction depending on the junction
length $d$ and and on the misorientation angle $\theta$. 
We show that the $\pi $ state disappears completely for $\theta > \frac{\pi}{2}$.
In the last section V, we discuss possible implications of our findings for
experimentally observed $0$--$\pi$ transitions in SFS junctions.

\section{Model}

\begin{figure}
    \includegraphics[width=8.6cm]{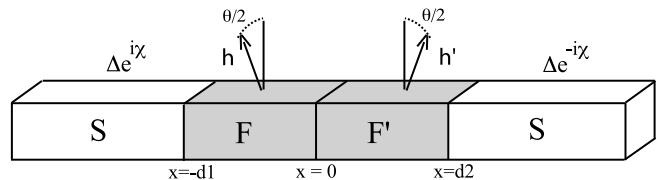}
    \caption{\label{sample} S-FF$'$-S junction with non-collinear magnetization.}
\end{figure}

We study a diffusive SFF$'$S Josephson junction with semi-infinite (that is, of
thickness much larger than $\xi_{S}$)
superconducting electrodes, as shown in Fig.\ \ref{sample}. 
The phase difference between the S-layers is denoted $\phi =2\chi $, the
thicknesses of the two ferromagnetic domains $d_{1}$ and $d_{2}$. 
In the following we consider a quasi-one-dimensional
geometry where the physical quantities do not depend on the in-plane
coordinates. For simplicity, we assume that the S-F and F-F$'$ interfaces are transparent.
We further assume that the temperature is close to $T_c$ so that $\Delta\ll T$, and
this allows us to linearize the Usadel equations.

In the case of superconducting--ferromagnet systems, the
proximity effect involves both the singlet and the triplet components
of the Green's functions.\cite{revbve} 
The Usadel equation in the ferromagnetic
layers takes the form (we follow the conventions 
used in Ref.\ \onlinecite{I+F})
\begin{equation}
D\nabla \left( \check{g}\nabla \check{g}\right) -\omega \left[ \hat{\tau}_{3}%
\hat{\sigma}_{0},\check{g}\right] -i\left[ \hat{\tau}_{3}\left( \mathbf{%
h\cdot \hat{\sigma}}\right) ,\check{g}\right] =0.  \label{usadel}
\end{equation}%
The Green function $\check{g}$ is a matrix in the Nambu $\otimes $ spin
space, $\hat{\tau}_{\alpha }$ and $\hat{\sigma}_{\alpha }$ denote the Pauli
matrices respectively in Nambu (particle-hole) and spin space, $\omega
=\left( 2n+1\right) \pi {T}$ are the Matsubara frequencies and $\mathbf{h}$
is the exchange field in the ferromagnet.  The Usadel equation is
supplemented with the normalization condition for the
quasiclassical Green function 
\begin{equation}
\check{g}^{2}={\check{\mathbf{1}}}=\hat{\tau}_{0}\hat{\sigma}_{0}.  \label{normalisation}
\end{equation}%
For simplicity, we assume that the superconductors are much less
disordered than the ferromagnets, and then we can impose the rigid
boundary conditions at the S/F interfaces:
\begin{equation}\label{bc}
\check{g}=\frac{1}{\sqrt{\omega^2+\Delta^2}}
\left( 
\begin{array}{cc}
\omega & \Delta e^{\pm{i}\chi } \\ 
-\Delta e^{\mp{i}\chi } & -\omega%
\end{array}%
\right)_{\rm Nambu} \otimes \hat{\sigma}_0\, ,
\end{equation}%
where $\Delta $ denotes the superconducting order parameter 
and the different signs refer respectively to the
boundary conditions at $x=-d_{1}$ and $x=d_{2}$. 

Close to the critical temperature $T_{c}$, the
superconducting correlations in the F region are weak,\cite{Rev} 
and we can linearize the Usadel equations (\ref{usadel}),(\ref{normalisation}) 
around the normal solution $\check{g}=\hat{\tau}_3\hat{\sigma}_0$.
The Green's function then takes the form
\begin{equation}
\check{g}=\left( 
\begin{array}{cc}
\sigma _{0}\sgn(\omega) & f_{\alpha }\sigma ^{\alpha } \\ 
-f_{\alpha }^{\dagger }\sigma ^{\alpha } & -\sigma _{0}\sgn(\omega)%
\end{array}%
\right) ,
\end{equation}%
where the scalar $f_{0}$ (respectively $f_{0}^{\dagger }$) and vector 
$\mathbf{f}$ (respectively $\mathbf{{f}^{\dagger }}$) components of the anomalous Green
functions obey the linear equations 
\begin{eqnarray}
\frac{\partial ^{2}f_{\pm}^{(\dagger )}}{\partial x^{2}}-\left[ \lambda _{\pm}%
\right] ^{2}f_{\pm}^{(\dagger )} &=&0  \notag \\
\frac{\partial ^{2}f_{\bot }^{(\dagger )}}{\partial x^{2}}-\left[ \lambda
_{\bot }\right] ^{2}f_{\bot }^{(\dagger )} &=&0 \label{wave equations}
\end{eqnarray}%
with
\begin{equation}
\lambda_{\pm }=\left[ 2\frac{\left\vert \omega \right\vert \mp
i h \sgn(\omega )}{D}\right] ^{1/2} , 
\qquad
\lambda _{\bot }=\left[ 2\frac{%
\left\vert \omega \right\vert }{D}\right] ^{1/2}.
\end{equation}
The projections of the anomalous Green function on the direction
of the exchange field (``parallel'' components) are defined as  $f_{\pm }^{(\dagger
)}(x)=f_{0}^{(\dagger )}\pm \mathbf{f}^{(\dagger )}\cdot \mathbf{e}_h$ where
$\mathbf{e}_h$ is the unit vector in the direction of the field.
The ``perpendicular'' component ${f}_{\bot }^{(\dagger )}$ refers to the axis orthogonal to the 
exchange field.
Generally, this component is a two-dimensional vector. In our
system, however, $\mathbf{f}$ lies in the plane spanned by the magnetizations
in the two domains, and therefore ${f}_{\bot }^{(\dagger )}$ has only
one component.

It follows from Eq.\ (\ref{wave equations}) 
that the decay of the ``parallel'' and the ``perpendicular'' components
is governed by two very different length scales. The parallel component
decays on the length scale  $\xi _{F}$, while the perpendicular component
is insensitive to the exchange field and decays on the typically
much larger scale $\xi _{S}=\xi_{F}\sqrt{\frac{h}{2\pi T_{c}}}$ 
(experimentally, $h$ may be more than 100 times larger than $T_c$, see, 
e.g., Ref.\ \onlinecite{Dead}).

In the absence of the exchange field, $f_{\sigma}$ and ${f}_{\sigma}^{\dagger }$
components are related by complex conjugation. The exchange field $h$ breaks
this symmetry, and the relation between $f_{\sigma}$ and 
${f}_{\sigma}^{\dagger}$ becomes
\begin{equation}
{f}_{\sigma}^{\dagger}(\chi) = f_{\sigma}(-\chi).
\label{f-symmetry}
\end{equation}

The solutions to the equations (\ref{wave equations}) in each of the ferromagnetic 
layers are given by
\begin{equation}
f_{\pm ,\bot }^{j}(x)=A_{\pm ,\bot }^{j}\sinh \lambda _{\pm ,\bot }x+B_{\pm
,\bot }^{j}\cosh \lambda _{\pm ,\bot }x,
\label{A-B-coefficients}
\end{equation}%
where the 12 coefficients $A_{\pm ,\bot }^{j}$ and $B_{\pm ,\bot }^{j}$ 
($j=1,2$ denotes the layer index) must be determined using the boundary
conditions at each interface. Note that it is enough to solve the equations
for the functions ${f}_{\sigma}^{j}$: the functions
${f}_{\sigma}^{j\dagger}$ can be then obtained from the symmetry
relation (\ref{f-symmetry})
 
As we assume transparent S/F interfaces and rigid boundary conditions,
\begin{eqnarray}
f_{\pm }^{1}(x=-d_{1}) &=&\frac{\Delta }{\omega }e^{i\chi }  \notag \\
f_{\pm }^{2}(x=d_{2}) &=&\frac{\Delta }{\omega }e^{-i\chi }  \notag \\
f_{\bot }^{1}(x=-d_{1}) &=&f_{\bot }^{2}(x=d_{2})=0\, .  \label{boundary1}
\end{eqnarray}%
At the (perfectly transparent) F/F$'$ interface, the standard 
Kupriyanov-Lukichev boundary conditions\cite{Lu} provide the 
continuity relations
\begin{eqnarray}
f_{\alpha }^{1}(x=0) &=&f_{\alpha }^{2}(x=0)  \notag \\
\frac{\partial f_{\alpha }^{1}}{\partial x}\vert_{x=0} 
&=&\frac{\partial f_{\alpha
}^{2}}{\partial x}\vert_{x=0}  \label{boundary2}
\end{eqnarray}%
(here $\alpha$ takes values 0 to 3 and refers to a fixed coordinate
system).
Note that, in the general case, since the
ferromagnetic exchange fields do not have the same orientation in the two
F-layers, the latter conditions do not lead to the continuity of the reduced
functions $f_{\pm ,\bot }^{j}$ and their derivative, except 
in the parallel case.

The last step will be to compute the Josephson current density 
using the formula\cite{revbve} 
\begin{equation}
I_{J}=ieN(0)DS\pi T\sum_{\omega =-\infty }^{\infty }\frac{1}{2}\mathrm{Tr}%
\left( \hat{\tau}_{3}\hat{\sigma}_{0}\check{g}\partial _{x}\check{g}\right),
\label{current}
\end{equation}%
where $S$ is the cross section of the junction, $N(0)$ is the density of 
states in the normal metal phase (per one spin direction) and the trace has to be
taken over Nambu and spin indices. The current can be explicitly 
rewritten for the linearized $\check{g}$
\begin{equation}\label{current2}
I_{J} = -ieN(0)DS\pi T\,\sum_{\omega =-\infty}^{\infty }\left[\sum_{\sigma=\pm}\frac{1}{2} (
f_{\sigma}\partial _{x}f_{\sigma}^{\dagger }-f_{\sigma}^{\dagger }\partial _{x}f_{\sigma})
+f_{\bot }\partial _{x}f_{\bot }^{\dagger }-f_{\bot }^{\dagger }\partial _{x}f_{\bot}
\right].
\end{equation}%
Using the coefficients introduced in
equations (\ref{A-B-coefficients}), the Josephson current (\ref{current2})
reads 
\begin{equation}\label{courant general}
I_{J}=ieN(0)DS\pi T\sum_{\omega, \sigma=\pm}\frac{\lambda _{\sigma }}{2}
\left[ A_{\sigma }(\chi )B_{\sigma }(-\chi)-B_{\sigma }(\chi )A_{\sigma }(-\chi)\right]
+\lambda _{\bot }
\left[ A_{\bot }(\chi )B_{\bot }(-\chi )- B_{\bot }(\chi )A_{\bot }(-\chi )\right]
.
\end{equation}%

Since the coefficients
$A^j_{\sigma }$ and $B^j_{\sigma }$ are solutions to the linear system
of equations (\ref{boundary1}) and (\ref{boundary2}), they are linear combinations
of $e^{i\chi}$ and $e^{-i\chi}$. Since the expression (\ref{courant general}) is
explicitly antisymmetric with respect to $\chi \mapsto -\chi$, it always
produces the sinusoidal current-phase relation (\ref{cpr}). Finally, the
expression (\ref{courant general}) does not contain the domain index $j$:
it can be calculated in any of the two domains, and the results must
coincide due to the conservation of the supercurrent in the Usadel equations.

In the following sections, 
this formalism is used to study the influence of a magnetic domain structure 
on the Josephson current.

\section{Domains of different thicknesses in the P and AP configurations}

\subsection{Parallel case (P)}

In the most trivial case $\theta=0$, the equations can be solved easily
with $A^j_{\bot }=B^j_{\bot }=0$. We naturally retrieve the expression 
reported in Ref.\ \onlinecite{Rev} for a single-domain S-F-S trilayer 
(of thickness $d_1+d_2$),
\begin{equation}
I^{P}_{c}=eN(0)DS\pi T\sum_{\omega,\sigma=\pm }\frac{\Delta^{2}
}{\omega ^{2}}\left[ \frac{\lambda _{\sigma}}{\sinh \lambda _{\sigma}(d_{1}+d_{2})}\right] 
\, .
\label{courant parallele somme}
\end{equation}%

The exact summation over the Matsubara frequencies $\omega$ can be done numerically. 
However, in many experimental situations, the exchange field is much larger than 
$T_{c}$. In this limit,
we can assume $h\gg\omega$ which implies
$\lambda _{\pm }=\frac{1 \mp i}{\xi_{F}}$. 
The summation over Matsubara frequencies reduces then to 
\begin{equation}
\sum_{\omega }\frac{1}{\omega ^{2}}=\frac{1}{4T^{2}}
\label{sum-Matsubara}
\end{equation}%
and the critical current is given by the simple expression 
\begin{equation}
I_{c}^{P}=I_{0}\mathrm{Re}\left[ \frac{1+i}{\sinh\left[(1+i)(\frac{d_{1}+d_{2}}{\xi_{F}})\right]}\right]
\label{courantcritiqueparallele}
\end{equation}%
with 
\begin{equation}
I_{0}=\frac{eN(0)DS\pi \Delta^2}{2\xi_{F} T}\, .
\end{equation}%
{}From Eq.\ (\ref{courantcritiqueparallele}) it is clear that the critical
current oscillates as a function of the junction length,
with a pseudo-period of the order of $\xi _{F}$. When the critical current
becomes negative, the S-F-S hybrid structure is in the $\pi $
state.

\subsection{Antiparallel case (AP)}

\begin{figure}
    \includegraphics[width=8.6cm]{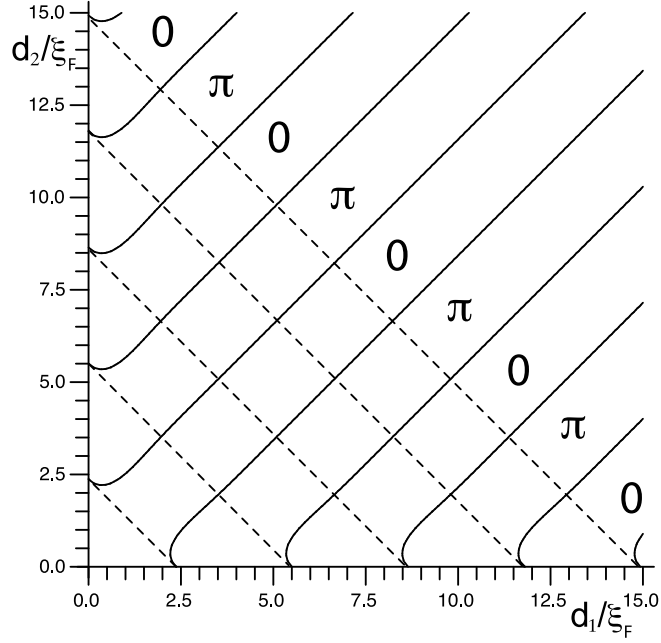}
    \caption{\label{AP} Quasi-periodic 0 to $\pi$ transitions for antiparallel (solid lines) and parallel (dashed lines) magnetization. On the graph, the indications 0 and $\pi$ refer to the antiparallel case. In the parallel case, the transitions occur along lines with $d_{1}+d_{2}=const$, starting from the 0 state.}
\end{figure}

In the antiparallel configuration $\theta=\pi$, the exchange
field has the opposite direction in the two domains. In this case we again find
$A^j_{\bot }=B^j_{\bot }=0$, and the critical current can be easily derived,
\begin{widetext}
\begin{equation}
I^{AP}_{c}=I_{0}\xi _{F}T^{2}\sum_{\omega ,\sigma =\pm }\frac{1}{\omega ^{2}}%
\left[ \frac{2\lambda _{\sigma }\lambda _{-\sigma }}{\lambda _{\sigma }\sinh
\lambda _{-\sigma }d_{2}\cosh \lambda _{\sigma}d_{1}+\lambda _{-\sigma }\cosh
\lambda _{-\sigma }d_{2}\sinh \lambda _{\sigma }d_{1}}\right]\, .
\label{courant antiparallele somme}
\end{equation}%
\end{widetext}
In the limit of the large exchange field $h\gg T_c$, the summation
over the Matsubara frequencies (\ref{sum-Matsubara}) results in 
\begin{equation}\label{courantcritiqueantiparallele}
I_{c}^{AP}=I_{0}\mathrm{Re}\left[\frac{2}{\sin(d_+ +i d_-)+\sinh(d_+ -id_-)} \right] 
\end{equation}%
with 
\begin{align}
d_+& =(d_{1}+d_{2})/\xi_F \\
d_-& =(d_{1}-d_{2})/\xi_F.
\end{align}%

For plotting the $0$--$\pi$ phase diagram in $d_{1}$--$d_2$ coordinates we use the
condition of the vanishing critical current.  From the 
equations (\ref{courantcritiqueparallele}) and (\ref{courantcritiqueantiparallele}), 
the critical current vanishes if 
\begin{equation}
\sin d_+\cosh d_+ +\sinh d_+ \cos d_+ =0  \label{annulationcasparallele}
\end{equation}%
in the parallel case, and if 
\begin{equation}
\sin d_+ \cosh d_- +\sinh d_+ \cos d_- =0
\label{annulationcasantiparallele}
\end{equation}%
in the antiparallel case. The resulting phase diagram is plotted in Fig.\ \ref{AP}.

For $d_{1}=d_{2}=d$ (symmetric case), we obtain that the critical current is positive for any
$d$: identical F layers in the AP configuration cannot produce the $\pi$ state
(a similar conclusion was drawn in Ref.\ \onlinecite{Hekking} for ballistic junctions
and for diffusive junctions at low temperature).
For $d_{1}\neq d_{2}$,  the
SFF$'$S junction can be either in the usual $0$ state or in the $\pi$ state depending on the difference between $d_{1}$ and $d_{2}$ (see Fig.\ \ref{AP}). 
For large $d_1$ and $d_2$, the periodic dependence of the phase transitions
on the layer thicknesses approximately corresponds to a single-layer SFS junction
of the thickness  $|d_1-d_2|+(\pi/4)\xi_F$. This result is similar to the case
of the clean SFF$'$S junction where the phase compensation arising from the two antiparallel
domains is observed.\cite{Hekking}

Another interesting feature of the phase diagram in Fig.\ \ref{AP} is the ``reentrant''
behavior of the phase transition at a very small thickness of one of the layers.
If the SFS junction is tuned to a $0$--$\pi$ transition point, and one adds a thin
layer F$'$ of antiparallel magnetization, then a small region of the ``opposite''
phase (corresponding to increasing the F thickness) 
appears, before the F--F$'$ compensation mechanism stabilizes
the phase corresponding to reducing the F thickness.

In this Section, we have seen that the $\pi$ state disappears 
in the antiparallel orientation for geometrically identical F-layers. However, we do not observe an enhancement of the critical current (compared to the zero field current) in the AP configuration such as reported in Refs.\ \onlinecite{Koshina2,enhancement}. This in agreement with the claim of Ref.\ \onlinecite{Koshina2} that this enhancement is present only at low temperatures.

In the next section, we demonstrate
that the suppression of the $\pi $ state occurs continuously 
as we change the misorientation angle.

\section{Arbitrary magnetization misorientation and equal thicknesses}

In the previous section, we have plotted the phase diagram for arbitrary layer thicknesses
$d_1$ and $d_2$ in the cases of parallel and antiparallel magnetization. In principle, one
can extend this phase diagram to arbitrary misorientation angles $\theta$. Such a calculation
amounts to solving a set of linear equations (\ref{boundary1}) and (\ref{boundary2})
for the 12 parameters defined in Eq.\ (\ref{A-B-coefficients}). This calculation is
straightforward, but cumbersome, and we consider only the simplest situation with
equal layer thicknesses $d_1=d_2=d$.

\begin{figure}
    \includegraphics[width=8.6cm]{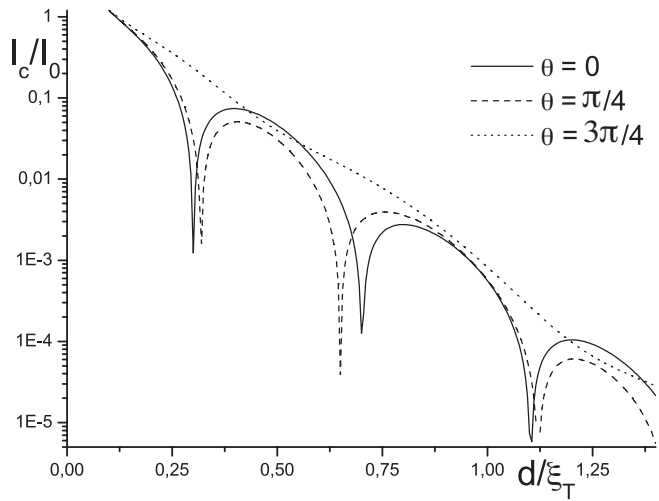}
    \caption{\label{exact} Critical current dependence on the size of the junction for $\theta=0,\pi/4\;\mathrm{and}\;3\,\pi/4$. We take $h/T=100$ which corresponds to a realistic value for a diluted ferromagnet, as reported in Ref.\ \onlinecite{Dead} and $\xi_{T}=\sqrt{D/2\pi{T}}$.}
\end{figure}

For equal layer thicknesses, the $0$--$\pi$ transitions are present at $\theta=0$ and
absent at $\theta=\pi$. We will see below that with increasing the 
misorientation angle $\theta$ the amplitude of the critical-current oscillations
(as a function of $d$) decreases, and the $\pi$ phase progressively shrinks.
At a certain ``critical'' angle $\theta_c$, the $\pi$ phase disappears completely
for any value of $d$. We find that the critical value is $\theta_{c}=\frac{\pi}{2}$, surprisingly independent of the strength of the exchange field.

The details of the calculation of the critical current are presented in the Appendix.
In the general case, the current can be written in the form of a Matsubara sum such 
as given in Eq.\ (\ref{courantcritiquethetasomme}). In Fig. \ref{exact}, we plot the 
current as a function of the domain thickness for different angles performing the 
summation over Matsubara frequencies numerically using realistic values for the 
temperature and the exchange field. We find that the domain structure reduces 
the $\pi$-state regions compared to the $\theta=0$ parallel case as well as the 
amplitude of the current in this state. To the contrary, the $0$-state regions 
are extended and the current amplitude is increased in this state. This result may be simply understood as a continuous interpolation
between a sign-changing $I_c$ in the single-domain case and an
always-positive $I_c$ in the antiparallel case.

\begin{figure}
    \includegraphics[width=8.6cm]{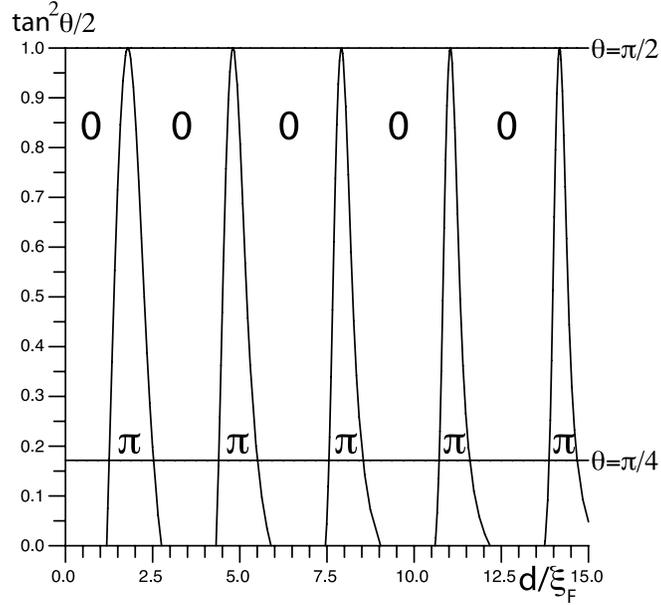}
    \caption{\label{gamma} $d$--$\theta$ phase diagram in the limit of a large exchange field. The dependence on $d$ is almost (but not exactly) periodic.}
\end{figure}

Considering the high-exchange-field limit introduced in Sec. III, namely 
$h\gg T_{c}$, and assuming further $d \ll \xi_S$ (which is a reasonable assumption for
the first several $0$--$\pi$ transitions in the high-field limit),
we have $\lambda _{\perp }\ll \lambda _{\pm
}$ and $\lambda _{\perp }d\ll 1$ so that one can expand 
Eq.\ \ref{courantcritiquethetasomme}) in powers of $\lambda _{\perp }$. To the lowest order
of expansion, the sum over Matsubara frequencies is done and we obtain
\begin{equation}
I_{c}(\theta )=8\frac{d}{\xi_{F}}I_{0}\frac{(Q_{+}+P_{+}\tan ^{2}\frac{\theta}{2} )(P_{+}+Q_{+}\tan
^{2}\frac{\theta}{2} )-(1-\tan ^{4}\frac{\theta}{2} )P_{-}Q_{-}}{(P_{+}^{2}-P_{-}^{2}+\tan
^{2}\frac{\theta}{2} (P_{+}Q_{+}+P_{-}Q_{-}))(Q_{+}^{2}-Q_{-}^{2}+\tan ^{2}\frac{\theta}{2}
(P_{+}Q_{+}+P_{-}Q_{-}))},  \label{courantcritiquetheta}
\end{equation}%
where $P_{\pm }$ and $Q_{\pm }$ are simple functions of the ratio $d/\xi_F$,  
\begin{align}
P_{\pm } &= 2\frac{d}{\xi _{F}}\left( \cosh (1+i)\frac{d}{\xi _{F}}\pm \cosh (1-i)%
\frac{d}{\xi _{F}}\right)  \notag \\
Q_{\pm } & =(1+i)\sinh (1-i)\frac{d}{\xi _{F}}\pm (1-i)\sinh (1+i)\frac{d}{\xi
_{F}}.  \label{PQ}
\end{align}%
{}From the general formula (\ref{courantcritiquetheta}), one can
retrieve the expressions 
(\ref{courantcritiqueparallele}) and (\ref{courantcritiqueantiparallele}) 
for the Josephson current in the (symmetric $d_{1}=d_{2}$) parallel and antiparallel cases
by setting respectively $\theta =0$ and $\theta =\pi$.
Within the approximation of a high exchange field, the critical current 
(\ref{courantcritiquetheta}) is a ratio of second degree polynomials 
in the variable $\tan ^{2}\frac{\theta}{2} $. The critical current cancels if 
\begin{equation}
\tan ^{4}\frac{\theta}{2} (P_{+}Q_{+}+P_{-}Q_{-})+\tan ^{2}\frac{\theta}{2}
(P_{+}^{2}+Q_{+}^{2})+(P_{+}Q_{+}-P_{-}Q_{-})=0.
\label{annulation courant theta}
\end{equation}
This equation allows one to compute the full S-FF'-S phase diagram in the $d$--$\theta$ 
coordinates (Fig.\ \ref{gamma}).
We observe that Eq.\ (\ref{annulation courant theta}) cannot be
satisfied for any thickness as soon as $\theta$ exceeds $\frac{\pi}{2}$. 
As the misorientation angle $\theta$ decreases below $\frac{\pi}{2}$,
the region of the $\pi $ state in the phase diagram Fig.\ \ref{gamma} increases, 
and it becomes maximal at $\theta =0$ (i.e., in the parallel
configuration). 

Away from the high-exchange-field limit, we can find the value of $\theta_{c}$ numerically using the exact formula, Eq.\ (\ref{courantcritiquethetasomme}). Remarkably, our calculations show that the critical value $\theta_c=\frac{\pi}{2}$ remains independent of the strength of the field $h$.

\section{Discussion and experimental aspects}

The main conclusion of the present work is that a domain structure in the SFS junction
reduces the region in the phase space occupied by the $\pi$ state. We have demonstrated
this reduction with the example of the two domains placed along the junction. However
we expect that this qualitative conclusion survives for more general configurations
of domains. In view of this reasoning, we suggest that the shift in the
$0$--$\pi$ transition sequence reported in Ref.\ \onlinecite{Dead} and attributed to
a ``dead layer'' of the ferromagnet may be alternatively explained by a domain
structure in the junction. If such a domain structure slightly reduces the region
of the $\pi$ phase in favor of the $0$ phase, this would shift the positions
of the two first $0$--$\pi$--$0$ transitions in a manner similar to the effect
of a ``dead layer'' (see, e.g., our $\theta=\pi/4$ plot in Fig.\ \ref{exact}). To
distinguish between the two scenarios, one would need to observe at least three
consecutive $0$--$\pi$ transitions (and/or develop a more realistic theory of
the effect of domains in SFS junctions).

Many of our results are based on the high-exchange-field approximation
assuming $\xi_S/\xi_F = \sqrt{h/(2\pi T_c)}\gg 1$. This is a reasonable
approximation for the type of samples reported in Ref.\ \onlinecite{Dead}:
the exchange field in the CuNi ferromagnetic alloys has
been estimated at 850K, whereas the critical temperature of Nb 
is of the order of 9K. Thus, the ratio $\xi_S/\xi_F$ is of the order of 4. Note that the high-field limit is consistent with the diffusive limit condition $h\tau_{e}\ll1$ with $\tau_{e}$ the elastic mean free time (see Ref.\ \onlinecite{Ko} for details). The parameters of the experiments\cite{Dead} yield the estimation $h\tau_{e}\approx0.1$.

In our treatment we have neglected the finite transparency
of the interfaces, the finite mean free path of electrons, the
spin-flip and spin-orbit scattering. Of course, those effects may be
incorporated in the formalism of Usadel equations in the usual way
(see, e.g., Refs.\ \onlinecite{Koshina,baladie,faure}). We expect that
they do not change the qualitative conclusion about the reduction
of the $\pi$ state by the domain structure. A realistic quantitative
theory of SFS junctions may need to take those effects into account,
in addition to a more realistic domain configuration in the ferromagnet.

\begin{acknowledgments}
We thank M.\ Houzet for useful discussions. This work was supported
by the Swiss National Foundation.
\end{acknowledgments}

\appendix

\section{Solving the linearized Usadel equations}

To solve the system of linear equations (\ref{boundary1}) and (\ref{boundary2})
with the 12 variables $A_{\pm ,\bot }^{j}$ and $B_{\pm ,\bot }^{j}$, it is convenient first
to reduce the number of variables by resolving the continuity relations (\ref{boundary2})
in terms of the 6 variables
\begin{align}
\beta _{\pm }& =B_{\pm }^{1}\mp B_{\perp }^{1}\tan \frac{\theta}{2}
=B_{\pm }^{2}\pm B_{\perp }^{2}\tan \frac{\theta}{2}  \notag \\
\beta _{\perp }& =B_{\perp }^{1}+\frac{B_{+}^{1}-B_{-}^{1}}{2}\tan \frac{\theta}{2}
=B_{\perp }^{2}-\frac{B_{+}^{2}-B_{-}^{2}}{2}\tan \frac{\theta}{2} \notag \\
\alpha _{\pm }& =\lambda_\pm A_{\pm }^{1} \mp \lambda_\perp A_{\perp }^{1}\tan \frac{\theta}{2}
=\lambda_\pm A_{\pm }^{2}\pm \lambda_\perp A_{\perp }^{2}\tan \frac{\theta}{2}  \\
\alpha _{\perp }& =\lambda_\perp A_{\perp }^{1}+
\frac{\lambda_+ A_{+}^{1}-\lambda_- A_{-}^{1}}{2}\tan \frac{\theta}{2}
=\lambda_\perp A_{\perp }^{2}-\frac{\lambda_+ A_{+}^{2}-\lambda_- A_{-}^{2}}{2}\tan \frac{\theta}{2}.
\notag
\label{beta}
\end{align}%
Solving now the set of 6 equations (\ref{boundary1}) produces the solution
\begin{align}
\alpha_{\perp} & =\frac{2\Delta}{\omega}\lambda
_{+}\lambda_{-}\lambda_{\perp}\cosh\left( \lambda_{\perp}d\right) \frac {p_{-}%
}{p_{+}^{2}-p_{-}^{2}+\tan^{2}\frac{\theta}{2}(p_{+}q_{+}+p_{-}q_{-})}\tan
\frac{\theta}{2}(1+\tan^{2}\frac{\theta}{2})\cos\chi \notag \\
\beta_{+}-\beta_{-} & =-\frac{4\Delta}{\omega}\lambda_{+}\lambda_{-}\sinh%
\left( \lambda_{\perp}d\right) \frac{p_{-}}{p_{+}^{2}-p_{-}^{2}+\tan
^{2}\frac{\theta}{2}(p_{+}q_{+}+p_{-}q_{-})}(1+\tan^{2}\frac{\theta}{2})\cos\chi \notag \\
\beta_{+}+\beta_{-} & =\frac{4\Delta}{\omega}\lambda_{+}\lambda_{-}\sinh%
\left( \lambda_{\perp}d\right) \frac{p_{+}+\tan^{2}\frac{\theta}{2} q_{+}}{%
p_{+}^{2}-p_{-}^{2}+\tan^{2}\frac{\theta}{2}(p_{+}q_{+}+p_{-}q_{-})}\cos\chi \notag \\
\alpha_{+}-\alpha_{-}& =-\frac{4i\Delta}{\omega }%
\lambda_{+}\lambda_{-}\lambda_{\perp}\cosh\left( \lambda_{\perp}d\right) 
\frac{q_{-}}{q_{+}^{2}-q_{-}^{2}+\tan^{2}\frac{\theta}{2}(p_{+}q_{+}+p_{-}q_{-})}(1+\tan^{2}\frac{\theta}{2})%
\sin\chi \\
\alpha_{+}+\alpha_{-} & =-\frac{4i\Delta}{\omega}\lambda_{+}\lambda
_{-}\lambda_{\perp}\cosh\left( \lambda_{\perp}d\right) \frac{q_{+}+\tan
^{2}\frac{\theta}{2} p_{+}}{q_{+}^{2}-q_{-}^{2}+\tan^{2}\frac{\theta}{2}(p_{+}q_{+}+p_{-}q_{-})}%
\sin\chi \notag \\
\beta_{\perp} & =\frac{2i\Delta}{\omega}\lambda_{+}\lambda_{-}\sinh\left(
\lambda_{\perp}d\right) \frac{q_{-}}{q_{+}^{2}-q_{-}^{2}+\tan^{2}\frac{\theta}{2}
(p_{+}q_{+}+p_{-}q_{-})}\tan\frac{\theta}{2}(1+\tan^{2}\frac{\theta}{2})\sin\chi \, ,
\notag
\end{align}
where
\begin{align}
p_{\pm }& =\lambda _{+}\lambda _{-}\sinh \lambda _{\perp }d\left( \cosh \lambda
_{+}d\pm \cosh \lambda _{-}d\right)
\notag \\
q_{\pm }& =\lambda _{\perp }\cosh \lambda _{\perp }d(\lambda _{+}\sinh \lambda
_{-}d\pm \lambda _{-}\sinh \lambda _{+}d) \, .
\end{align}%

In terms of the new variables, the supercurrent (\ref{courant general}) becomes
\begin{align}
I_{J}& =ieN(0)DS\pi T \, \sum_{\omega }\left[ \frac{1}{4%
}(\alpha _{+}+\alpha _{-})(\chi )(\beta _{+}+\beta _{-})(-\chi )\right.  \notag \\
& \left. +\frac{1}{4(1+\tan ^{2}\frac{\theta}{2} )}(\alpha _{+}-\alpha _{-})(\chi
)(\beta _{+}-\beta _{-})(-\chi )+\frac{\alpha _{\perp }(\chi )\beta _{\perp
}(-\chi )}{1+\tan ^{2}\frac{\theta}{2} }\right]-\left[\chi\leftrightarrow-\chi\right].
\end{align}
The resulting current-phase relation is sinusoidal with the critical current
\begin{align}
I_{c}(\theta )& =4I_{0}\xi _{F}T^{2}\sum_{\omega }\frac{(\lambda _{+}\lambda
_{-})^{2}\lambda _{\perp }\sinh 2\lambda _{\perp }d}{\omega ^{2}}
\label{courantcritiquethetasomme} \\
& \times \frac{(q_{+}+p_{+}\tan ^{2}\frac{\theta}{2} )(p_{+}+q_{+}\tan ^{2}\frac{\theta}{2}
)-(1-\tan ^{4}\frac{\theta}{2} )p_{-}q_{-}}{(p_{+}^{2}-p_{-}^{2}+\tan ^{2}\frac{\theta}{2}
(p_{+}q_{+}+p_{-}q_{-}))(q_{+}^{2}-q_{-}^{2}+\tan ^{2}\frac{\theta}{2}
(p_{+}q_{+}+p_{-}q_{-}))} . \notag
\end{align}%
Eq.\ (\ref{courantcritiquethetasomme}) can be used for numerical calculations of the
critical current for an arbitrary relative orientation of the ferromagnetic exchange fields 
and for any value of their magnitude (e.g., for producing the plot in Fig.\ \ref{exact}). 
In the body of the article, a simpler expression for the current is given 
in the high-exchange-field limit (Eq.\ (\ref{courantcritiquetheta})).

\bibliography{SFS2}

\end{document}